\documentstyle[12pt,twoside,fleqn,espcrc1,epsfig]{article}

\newcommand{\AmS}{{\protect\the\textfont2
  A\kern-.1667em\lower.5ex\hbox{M}\kern-.125emS}}

\title{Effect of $\sigma$(600)-Production 
in $p\bar p\rightarrow 3\pi^0$ at rest}

\author{M. Y. Ishida\address{Department of Physics, 
Tokyo Institute of Technology,Tokyo 152-8551, Japan}, 
T. Komada\address{Coll. Sci. and Tech. Nihon U., 
Kanda-Surugadai, Chiyoda, Tokyo 101-0062, Japan}, 
S. Ishida$^{\rm b}$,
T. Ishida\address{KEK, Oho, Tsukuba, Ibaraki 305-0801, Japan},
K. Takamatsu\address{Miyazaki U., Gakuen-Kibanadai, 
Miyazaki 889-2155, Japan} and
T. Tsuru$^{\rm c}$}
       
\begin{document}

% typeset front matter
\maketitle

\begin{abstract}
The $\pi^0\pi^0$ mass spectra and angular distributions
around $K\bar K$-threshold and at 1.5 GeV in 
$p\bar p({\rm at\ rest})\rightarrow 3\pi^0$  
in the Crystal Barrel experiment are reanalyzed by applying
the new method, which is consistent with unitarity of $S$-matrix
and expressed directly by resonance parameters.
The effects of light $\sigma$-meson production are clearly seen
to improve the fit with $\sigma$, 
in comparing with the fit without $\sigma$.  
\end{abstract}

\section{Introduction}
The light iso-singlet scalar $\sigma$ meson plays an important role
in the mechanism of spontaneous breaking of chiral symmetry, 
and to confirm its real existence is
one of the most important topics in hadron physics.
Recently in various $\pi\pi$-production experiments\cite{rf2,rf4} 
a broad peak 
is observed in mass spectra below 1 GeV.
Conventionally this peak was regarded as a mere non-resonant 
background, basing on the ``universality argument,''\cite{rf3} since
no $\sigma$ was seen in $\pi\pi$ scatering at that time.
However, at present the $\pi\pi$-scattering phase shift $\delta_S^{I=0}$
 is reanalyzed
by many authors\cite{rf5,rf1,rf2} and the existence of $\sigma$ meson is 
strongly suggested. 
A reason of missing $\sigma$ in the 
conventional analysis is pointed out to be due to 
overlooking the cancellation mechanism,\cite{rf2} which is guaranteed 
by chiral symmetry, between the effects of $\sigma$ and 
of repulsive $\pi\pi$-interaction.
Moreover, the conventional treatment, based on the universality argument,
of the low mass broad peak was shown to be not correct, and
a new effective (VMW) method is proposed to analyze 
resonance productions.
In this method the production amplitude ${\cal F}$ is 
directly represented by
the sum of Breit-Wigner amplitudes with production couplings
and phase factors (, including initial 
strong phases) of relevant resonances.
The consistency of this method with 
the unitarity is seen from the following field 
theoretical viewpoint\cite{rf2}.

Presently, after knowing the quark physics, the strong interaction
 ${\cal L}_{\rm str}$ among hadrons (in our example, mesons) is regarded
as a residual interaction of QCD among 
color singlet $q\bar q$-bound states, the ``bare states,'' denoted as
$\pi =\bar\pi ,\bar\sigma ,\bar f_0$ and $\bar f_2$.
In switching off  
${\cal L}_{\rm str}$, the bare states appear as stable particles 
with zero widths. In switching on ${\cal L}_{\rm str}$,
they change into the physical states
%(, denoted as $\pi ,\sigma , f_0$ and $f_2$,) 
with finite widths.
The unitarity of $S$-matrix is guaranteed automatically
by the hermiticity
of ${\cal L}_{\rm str}$.  
The VMW method is obtained directly as the representation of  
production amplitude by physical state bases,
with a diagonal mass and width. 

The VMW method is already applied\cite{rf2} to the 
analyses of $pp$-central collision
$pp\rightarrow pp\pi^0\pi^0$ and $J/\psi\rightarrow\omega\pi\pi$ decay,
leading to a strong evidence of existence of the light $\sigma$ meson.
In this talk we apply this method to the analysis of 
$p\bar p\rightarrow 3\pi^0$ at rest.

\section{Amplitude describing $p\bar p\rightarrow 3\pi^0$ at rest
by VMW method}
\subsection{ ${\cal L}_{\rm str}$
for $p\bar p\rightarrow 3\pi^0$ at rest}
We apply the iso-bar model, describing the process in following two steps:
In the first step the $p\bar p$ annihilates into the resonance 
$f(f_0{\rm \ and\ }f_2)$ and $\pi^0$, and 
the $f$ decays into $2\pi^0$ in the second step.
Since 
both $p$ and $\bar p$ are at rest,
the relative momentum $p_\mu =p_{p\mu}-p_{\bar p\mu}$
(and the relative angular momentum $L_{p\bar p}$) between 
$p$ and $\bar p$ is 0. Charge conjugation parity $P_C$ of $f\pi^0$-system 
is +1. Thus the three types of $(\bar p,p)$ bi-linear forms
%\footnote{This implies 
%the initial $p\bar p$ is in 
%$^1S_0$ state. In the original analysis\cite{rf4}, 
%the phenomenological parameters related with those  
%$^3P_1$ and $^3P_2$ states not allowed theoretically 
%are included.
%}
with $P_C$=+1 are possible:
$\bar pi\gamma_5p,\bar pi\gamma_5\gamma_\mu p$ and $\bar pp$.
The second type reduces to the first type by using the 
equation $-\partial_\mu (\bar pi\gamma_5p)=2m_p\bar pi\gamma_5\gamma_\mu p
+\bar pi\gamma_5i\sigma_{\mu\nu}
\stackrel{\leftrightarrow}{\partial_\nu} p
=2m_p\bar pi\gamma_5\gamma_\mu p$ (, derived from 
Dirac equation 
and the above ``rest'' condition),
and the third type is forbidden by parity. Thus only the first type
remains.\footnote{This implies 
initial $p\bar p$ is in 
$^1S_0$ state. In the original analysis\cite{rf4}, 
the phenomenological parameters related with 
$^3P_1$ and $^3P_2$ states, considered not to contribute
since of the above rest condition,  
are included.
}
The most simple form of
 ${\cal L}_{\rm str}^{1,2}$ describing the 1st step 
$p\bar p\rightarrow f\pi^0$ 
and the 2nd step $f\rightarrow 2\pi^0$ is given, respectively, 
by

\vspace*{-0.5cm}

\begin{eqnarray}
{\cal L}_{\rm str}^1 = 
\sum_{\bar f_0,\bar f_2}(\bar\xi_{\bar f_0}\bar pi\gamma_5p\bar f_0\pi^0
+\bar\xi_{\bar f_2}\bar pi\gamma_5p\bar f_{2\mu\nu}
\partial_\mu\partial_\nu\pi^0), \  
{\cal L}_{\rm str}^2 =
\sum_{\bar f_0,\bar f_2}(\bar g_{\bar f_0}\bar f_0\pi^2
+\bar g_{\bar f_2}\bar f_{2\mu\nu}
(\pi\stackrel{\leftrightarrow}{\partial_\mu}
\stackrel{\leftrightarrow}{\partial_\nu}\pi )).
\nonumber
\end{eqnarray}
\vspace*{-0.5cm}

\subsection{Amplitude by VMW method}
First
denoting the three $\pi^0$ as $\pi_1,\pi_2$ and $\pi_3$
with momenta $p_1,p_2$ and $p_3$, respectively, we   
consider the $\pi_1$ and $\pi_2$ forming the 
resonance $f$ with 
squared mass $s_{12}$
(where $s_{ij}\equiv -(p_i+p_j)^2$)
% and 
%$s_{12}+s_{23}+s_{13}=4m_p^2+3m_\pi^2$)
 and  with momentum 
$|{\bf p}|$  in z-direction. 
The $\pi_1$ has
momentum $|{\bf q}|$ and polar angle $\theta$ 
in the $f$ rest frame.
In the lowest order in bare state representation the
amplitude is $2im_pf^{s_ps_{\bar p}}
(\sum_{\bar f_0}
\frac{\bar\xi_{\bar f_0}\bar g_{\bar f_0}}{\bar m_{\bar f_0}^2-s_{12}}
+\sum_{\bar f_2}
\frac{\bar\xi_{\bar f_2}\bar g_{\bar f_2}N(s_{12},{\rm cos} \theta_{12})}
{\bar m_{\bar f_2}^2-s_{12}})$,
where $N(s_{12},{\rm cos}\theta_{12})=-\frac{(s_{23}-s_{31})^2}{4}
+\frac{16m_p^2{\bf p}^2{\bf q}^2}{3s_{12}}$,\footnote{
$N$ is obtained by the calculation of $N=-p_{3\mu}p_{3\nu}
{\cal P}_{\mu\nu ;\lambda\kappa}
(p_1-p_2)_\lambda (p_1-p_2)_\kappa$, 
where we use the tensor projection operator 
${\cal P}_{\mu\nu ;\lambda\kappa}$ with mass squared $s_{12}$
instead of $m_{f_2}^2$.
}
and 
$2im_pf^{s_ps_{\bar p}}\equiv\bar p({\bf 0},s_{\bar p})i\gamma_5
p({\bf 0},s_p)$($s_{\bar p}$ and $s_p$ being spin of $\bar p$ and $p$, 
respectively, and $f^{++}=f^{--}=0,f^{+-}=-f^{-+}=-1$).
Owing to the effect of final (and initial) state interaction, 
the ``full order'' of the amplitude 
in physical state representation is given by
\vspace*{-0.5cm}
\begin{eqnarray}
A_{s_ps_{\bar p} }(s_{12},{\rm cos} \theta_{12})
=2im_pf^{s_ps_{\bar p} }
\left( \sum_{f_0}
\frac{r_{f_0}e^{i\theta_{f_0}}}{m_{f_0}^2-s_{12}-i\sqrt{s_{12}}
\Gamma_{f_0}(s_{12})} 
+  \sum_{f_2}
\frac{r_{f_2}e^{i\theta_{f_2}}N(s_{12},{\rm cos} \theta_{12})}
{m_{f_2}^2-s_{12}-i\sqrt{s_{12}}
\Gamma_{f_2}}        \right) .\nonumber
\end{eqnarray}
The symmetric amplitude 
${\cal F}_{s_ps_{\bar p}}$ is obtained simply 
by its cyclic sum as\\
%\begin{eqnarray} 
$
{\cal F}_{s_ps_{\bar p}}(s_{12},s_{23},s_{31}) =
A_{s_ps_{\bar p}}(s_{12},{\rm cos} \theta_{12})
+A_{s_ps_{\bar p}}(s_{23},{\rm cos} \theta_{23})
+A_{s_ps_{\bar p}}(s_{31},{\rm cos} \theta_{31}).
%\end{eqnarray}
$\\
Cross section is given by 
%\begin{eqnarray} 
$
d\sigma\sim\int_{2m_\pi}^{2m_p-m_\pi}d\sqrt{s_{12}}
\frac{\sqrt{s_{12}}}{\pi}\frac{|{\bf p}|}{8\pi m_p}
\frac{|{\bf q}|}{8\pi\sqrt{s_{12}}}\int_{-1}^1d{\rm cos}\theta
\overline{|{\cal F}(s_{12},s_{23},s_{31})|^2},
%\end{eqnarray}
$
where $\overline{|{\cal F}(s_{12},s_{23},s_{31})|^2}
\equiv (1/4)\sum_{s_p,s_{\bar p}}|{\cal F}_{s_ps_{\bar p}}(s_{12},s_{23},s_{31})|^2
=(1/2)|{\cal F}_{+-}(s_{12},s_{23},s_{31})|^2$, and
$s_{12}+s_{23}+s_{13}=4m_p^2+3m_\pi^2$,
$s_{23}=-4m_p(|{\bf p}||{\bf q}|/\sqrt{s_{12}}){\rm cos}\theta
+m_\pi^2+2m_pE_3;\ 
E_3=\sqrt{m_\pi^2+{\bf p}^2}=(4m_p^2-s_{12}+m_\pi^2)/4m_p.$

\section{Results}
We use as experimental data the $\pi^0\pi^0$ mass spectra and angular 
distributions around $K\bar K$-threshold and at 1.5 GeV, 
which are published
in the paper\cite{rf4} by Crystal Barrel collaboration. We take into 
consideration as the physical particles   
$f_0=\sigma ,f_0(980), f_0(1370)$, $f_0(1500)$,
and $f_2=f_2(1275)$,$f_2(1565)$.
Preliminary result of our fit is shown in Fig. 1. The mass and width of 
$\sigma$ obtained are  
%%%%%%%%%%%%%%%%%%%%%%%%%%%%%%%%%%%%%%%%%%%%
$m_\sigma =560\pm 25$MeV and 
$\Gamma_\sigma =350\pm 50$MeV
(error corresponding to the 5$\sigma$-deviation),
and the $\chi^2$ is given by $858/(282-25)=3.3$.
%%%%%%%%%%%%%%%%%%%%%%%%%%%%%%%%%%%%%%%%%%%%
The spectra given by 
setting $r_\sigma =0$
are also given by dashed lines. Effect of $\sigma$-production is 
seen to be crucially important in reproducing the structure of
mass spectra below 1 GeV.

\begin{figure}[t]
  \epsfysize=8 cm
 \centerline{\epsffile{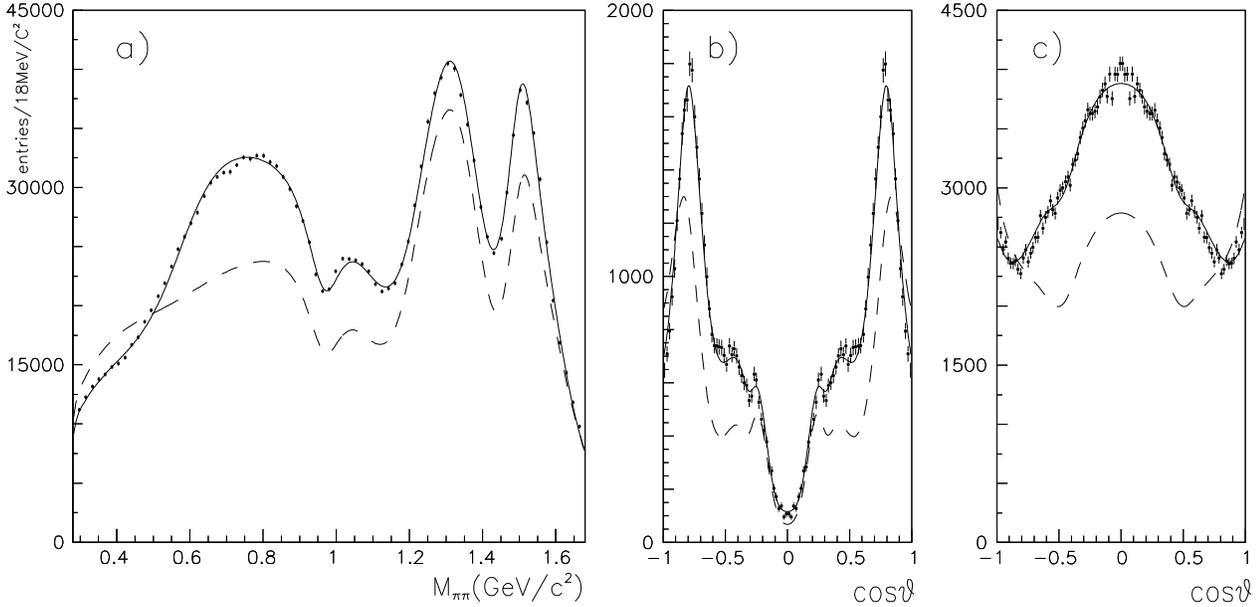}}
 \caption{ Result of the fit:
(a) $\pi^0\pi^0$ mass distribution, and 
cos$\theta$ distributions (b) around $K\bar K$-threshold and 
(c) at 1.5 GeV.
Dashed lines represent the spectra obtained by setting 
$r_\sigma =0.$ }
  \label{fig:2}
\end{figure}

%Property of $\sigma$ is somewhat uncertain, depending on fit A and fit B,
%since the effect of $\sigma$ is difficult to be discriminated from that
%of $f_0(1370)$ due to the statistics of $3\pi^0$. In fit A the
%$f_0(1370)$ has wide width about 600 MeV, and its effect is much larger 
%than $\sigma$-effect. In fit B the $f_0(1370)$ 
%has rather a narrow width about 200 MeV, and its effect is comparable 
%order of $\sigma$-effect.

We have also tried to fit without introducing the $\sigma$ Breit-Wigner
amplitude. 
%by setting $r_\sigma =0$. 
%%%%%%%%%%%%%%%%%%%%%%%%%%%%%%%%%%%%%%%%%%%%%%%%%%%%
The corresponding $\chi^2$ is 
3465/(282-21)=13.3,
%%%%%%%%%%%%%%%%%%%%%%%%%%%%%%%%%%%%%%%%%%%%%%%%%%%%
which is much worse than our best fit with
$\sigma$ meson. 
This gives a strong evidence for $\sigma$-existence.

\section{Comparison of the methods of analyses between CB's and Ours}
In the original analysis by Crystal Barrel collaboration
the $\pi\pi$ scattering and production amplitudes are
given, respectively, as in 
the ${\cal K}$ matrix representation:
${\cal T}={\cal K}/(1-i\rho{\cal K}),\ \ 
{\cal F}={\cal P}/(1-i\rho{\cal K})$, where ${\cal K}$-matrix
and ${\cal P}$-matrix are taken 
in pole dominative form;
%\begin{eqnarray}
$
{\cal K} = \sum_\alpha g_\alpha^2/(m_\alpha^2-s)+c_{\rm BG},\ \ 
{\cal P} = \sum_\alpha
e^{i\theta_\alpha}\xi_\alpha g_\alpha /(m_\alpha^2-s).
$
%\label{eq7}
%\end{eqnarray}
Here 
the summation is taken for the ${\cal K}$-matrix states
$\alpha$, which are related 
%through the complex orthogonal transformation 
to the physical states corresponding to the poles of 
${\cal T}$ (or ${\cal F}$).
In the case with no production phases, $\theta_\alpha =0$,
%${\cal P}$ becoming real, 
the ${\cal F}$ and ${\cal T}$
have the same phase, 
%coming from the common factor $1/(1-i\rho{\cal K})$,
and the unitarity(final state interaction theorem) is satisfied.
In their analysis this phase was set to be the experimental scattering
phase shift  
$\delta_{\rm S}^{\rm I=0}$.
However, in the above ${\cal K}$-matrix parametrization, 
%of Eq.(\ref{eq7}), 
when $s$ is close to $m_\alpha^2$, ${\cal K}$
diverges and the phase must take the
value $90^\circ (+n$$\times$$180^\circ)$.
This gives a very strong constraint for the value of $m_\alpha$.
The experimental $\delta_{\rm S}^{\rm I=0}$ passes through
$90^\circ$ at about $\sqrt{s}\simeq 900$ MeV, and so the $m_\alpha$ 
becomes $m_\alpha\simeq 900$ MeV ,
which is much larger than $m_\sigma$(=600MeV).
Thus, no existence of light $\sigma$ is implicitly assumed 
from the beginning. 

In our method, the $\delta_{\rm S}^{\rm I=0}$
is analyzed by introducing the repulsive
background phase shift $\delta_{\rm BG}$, which is required from   
chiral symmetry.\cite{rf2}
\footnote{
In our method the $S$-matrix is  
parametrized by $S=S^{\rm Res}S^{\rm BG}$ and the ${\cal K}$-matrix
by ${\cal K}=\frac{{\cal K}^{\rm Res}+{\cal K}_{\rm BG}}
{1-\rho^2{\cal K}^{\rm Res}{\cal K}_{\rm BG}}$. 
The denominator removes the poles of ${\cal K}^{\rm Res}=\sum_\alpha
g_\alpha^2/(m_\alpha^2-s)$
in the total ${\cal K}$-matrix and we can take the small 
$m_\alpha\simeq 600$MeV. On the other hand, 
the background matrix, $c_{BG}$,
in the conventional ${\cal K}$-matrix
cannot describe 
the global phase motion corresponding to $\delta_{\rm BG}$ in our method, 
and the small value of $m_\alpha$
is not permissible.}
%$=\delta_\sigma +\delta_{\rm BG}+\delta^{\rm Res}_{\rm others}$.
Correspondingly, the 
${\cal F}$ is represented by 
${\cal F}=\frac{{\cal P}^{\rm Res}}{1-i\rho{\cal K}^{\rm Res}}
e^{i\delta_{\rm BG}}$,\footnote{Here we neglect the possible effect of
non-resonant $3\pi^0$ production. } 
which satisfies the final state interaction theorem.
% since 
%Arg $\frac{e^{i\delta_{\rm BG}}}{1-i\rho{\cal K}^{\rm Res}}
%=\delta^{\rm Res}+\delta_{\rm BG}=\delta_{\rm S}^{\rm I=0}$.
This ${\cal F}$ is rewritten into the form, applied in VMW method,
in the physical state representation.
%${\cal F}=(\sum_fr_fe^{i\theta_f}
%\Delta_f^{\rm BW}(s))e^{i\delta_{\rm BG}}(,\ \Delta_f^{\rm BW}(s)$ 
%being the Breit-Wigner amplitude of resonance $f$),
%which is the production amplitude 
In our approach,
whether $\sigma$ meson exists or not is determined directly from
the experimental data themselves, as was done in \S 3.

\section{Conclusion}
Through the above results
the effects of production of light $\sigma (600)$ meson 
are clearly shown, while in 
the original analysis by Crystal Barrel collaboration,
no existence of light $\sigma$ is implicitly assumed 
from the beginning.
Mass and width of $\sigma$
is obtained as 
%%%%%%%%%%%%%%%%%%%%%%%%%%%%%%%%%%%%%%%%%%%%%%%
$m_\sigma =560\pm 25$MeV, 
$\Gamma_\sigma =350\pm 50$MeV, 
%%%%%%%%%%%%%%%%%%%%%%%%%%%%%%%%%%%%%%%%%%%%%%%
which are consistent with those 
obtained in our previous phase shift analysis\cite{rf1}
($(m_\sigma ,\Gamma_\sigma )$=$(535\sim 675, 385\pm 70)$MeV).
However, the effect of $\sigma (600)$ in this process 
is difficult to be discriminated from that of $f_0(1370)$
due to 
the statistics property of $3\pi^0$ system. 
In order to avoid this, 
it is desirable to analyze also the process, 
$\bar pn\rightarrow \pi^0\pi^0\pi^-$,
through the similar method.

\end{document}